# Kirchhoff's Current Law Can Be Exact

*available at https://arxiv.org/abs/1905.13574*

Robert S. Eisenberg
Department of Applied Mathematics
Illinois Institute of Technology;

Department of Physiology and Biophysics
Rush University Medical Center
Chicago IL

USA

Bob.Eisenberg@gmail.com

December 5, 2022

*File name: Kirchhoff's Current Law Can Be Exact with Addendum December 5-2 2022.docx*

**Addendum with more Recent References**

*than found in the version arXiv:1905.13574v3, 18 Jul 2019*

December 5, 2022

**Bibliographical Note**.

The December 5, 2022 version of this paper includes
the revision *arXiv:1905.13574v3, 18 Jul 2019* without changes.
It also contains the addition "**More Recent Work**".

Sans serif font [Calibri] is used for the Addendum
so it is unlikely to be confused with *arXiv:1905.13574v3, 18 Jul 2019*

References in the two parts of this paper are distinct and independent
and have different format.
References in the component *arXiv:1905.13574v3, 18 Jul 2019* are not changed
from the original publication and have their original formatting.
They are shown in the original serif font, Times New Roman.

Page numbering is also distinct and independent in the two sections,
made clear by the placement of page numbers in the two sections,
as well as the different fonts,
I hope.

## More recent work

A number of papers [1-13] have been written by my collaborators and me about Kirchhoff's current law since the version *arXiv:1905.13574v3, 18 Jul 2019* of this paper. These more recent papers report my journey, my odyssey to a safe intellectual home.

I worked to understand why Kirchhoff's current law is so widely used on time scales far outside the times at which the law is derived in textbooks. Kirchhoff's current law is used as the main design tool for circuits that operate in $10^{-10}$sec although it is presented as a long time approximation (say $> 10^{-3}$ sec) in textbooks and Wikipedia and derived that way in the literature.

The derivations also do not deal with the obvious fact that $\mathbf{J}$ (the movement of charge with mass) does not follow a conservation law at all, as is obvious from the Maxwell-Ampere law, eq. (1). In fact, $\mathbf{J}$ obeys the continuity equation, $\mathbf{div}\,\mathbf{J} = -\mathbf{div}\,(\varepsilon_0\ \partial\mathbf{E}/\partial t)$, not the zero value the divergence would have if $\mathbf{J}$ were conserved. This difficulty seems not to have been discussed in textbooks or elsewhere (Compare with the worked example, reference [25] in the arXiv:1905.13574v3, 18 Jul 2019). The difficulty actually shows that Kirchhoff's law as usually presented is not an approximation to the electrodynamic Maxwell equations, except at DC, and the equations are not dynamic at all, because $\partial\mathbf{E}/\partial t = 0$. Under other conditions, ***<u>Kirchhoff's law</u>*** as discussed in textbooks is ***<u>incompatible with the Maxwell equations</u>***, a fact that might be of interest to students of those textbooks.

**<u>A 'final' word</u>.** Reference [2] attempts to be a 'final word' about deriving Kirchhoff's law, because ***it gives the necessary conditions for Kirchhoff's law*** for current flow in circuits. Reference [2] tries to show what must be present in Kirchhoff's law if it is consistent with the Maxwell equations.

Reference [2] does ***not*** try to give definitive ***sufficient conditions*** for Kirchhoff's law in real circuits. No final word seems to be possible for sufficient conditions. Every real circuit is likely to be a special case.

Reference [2] attempts to state sufficient conditions for a different class of circuits, namely the simplified circuits of elementary electrical circuit theory. Those simplified circuits satisfy Kirchhoff's law ***if*** current is generalized to include the displacement current $\varepsilon_0\ \partial\mathbf{E}/\partial t$, as described in *arXiv:1905.13574v3, 18 Jul 2019*. More specifically, a conjecture is made that the redefinition (eq. (3)) of current in Kirchhoff's law as $\mathbf{J}_{total} = \mathbf{J} + \varepsilon_0\ \partial\mathbf{E}/\partial t$ is ***both*** necessary and sufficient but only ***in the simplified circuits*** of elementary over-simplified electrical circuit theory. See the worked example, reference [25] in the *arXiv:1905.13574v3, 18 Jul 2019*. These over-simplified circuits do not pretend to describe all the properties of real circuits. Rather over-simplified circuits are the foundation on which design is based, with more realistic treatments of the physics of components, layout, and coupling added as needed for the application of interest. Simplified circuits do not include realistic descriptions of wires or other circuit components, mutual inductance, and other complexities, even stray capacitances.

More recent papers from other research groups are not included here, because they have not been carefully studied. Other work distracted me [13,16-31] from a careful reading of the developing literature. Apologies are extended to those whose contributions have not been recognized, however inadvertently.

**Displacement current** $\varepsilon_0\ \partial\mathbf{E}/\partial t$ is called the 'ethereal current' in many of my papers to emphasize how closely it is related to the properties of the aether (UK spelling; USA spelling is sometimes 'ether').

The ethereal current $\varepsilon_0\ \partial\mathbf{E}/\partial t$ is produced by the 'polarization of space' as Maxwell called it (p. 416 of [14]). It exists everywhere, in the vacuum of outer space, as well as in the space between atoms, and inside the atoms themselves.

Some have called $\varepsilon_0\ \partial\mathbf{E}/\partial t$ the 'Maxwell current' but I prefer to reserve that name as an alternative term for the ***entire*** source term for $\mathbf{curl\ B}$, i.e., the entire right hand side of the Maxwell Ampere law eq. (1), $\mathbf{J}_{total} = \mathbf{J} + \varepsilon_0\ \partial\mathbf{E}/\partial t$.

**What form of electricity flows in a vacuum?** The word 'flow'—in ordinary English/American—implies that something flows. So does the word 'current'. (Think of the current in a river, as an example.) Flow and current are emergent phenomena, but the abstraction of flow emerges from something concrete, in the normal use of the word 'flow'. Flow and current describe an underlying phenomenon in which something moves in ordinary language..

In particular, the ordinary use of language implies that if current $\varepsilon_0\ \partial\mathbf{E}/\partial t$ exists in a vacuum, something flows to carry the current in that vacuum, even in a vacuum devoid of mass. Maxwell certainly shared this view of the meaning of $\varepsilon_0\ \partial\mathbf{E}/\partial t$ . He spent many years trying to build a mechanical model for $\varepsilon_0\ \partial\mathbf{E}/\partial t$ and the (movement of the) aether that created the flow.

But physics requires more than the ordinary use of language. It requires a specific definition of what might flow in a vacuum. It requires a way to measure what might flow, at least in principle. It requires a mathematical description of what might flow in the vacuum. The mathematical description of the flow is known, namely $\varepsilon_0\ \partial\mathbf{E}/\partial t$. If one wants to define an aether that flows to create $\varepsilon_0\ \partial\mathbf{E}/\partial t$, the properties of the aether itself must be defined precisely.

In the traditions of modern physics, the aether that might flow in a vacuum could follow rules much more general than the mechanical models considered by Maxwell. Quasi-particles like holes and electrons have been defined abstractly and have allowed the development of solid state devices of technology and all that depends on them. These quasi-particles follow the same Poisson Drift Diffusion equations that real ions follow as described in biophysics and chemistry by the PNP (Poisson Nernst Planck) equations.

It remains unclear what (if anything) supports current flow in a vacuum. ***Could a quasi-particle be invented that carries the ethereal current $\varepsilon_0\ \partial E/\partial t$, even in a vacuum?*** It is not clear that such an ethereal quasi charge can be sensibly defined without violating the laws of physics or mathematics. It is clear that electrodynamics has done well without that definition.

**Historical context**. Reference [1] describes the historical context in which current in Kirchhoff's law was defined. David Ferry found a paper by Kirchhoff that demonstrates Kirchhoff's understanding of total current, and the need to include $\varepsilon_0\ \partial\mathbf{E}/\partial t$.

Reference [1] also describes the modern perspective. That paper discusses the quantum mechanical use of Kirchhoff's law for total current in the actual design of modern high speed computing circuits, a subject that two of the authors (DF & XO) have worked on for many years. Modern high speed circuits switch in $10^{-10}$ seconds and are in fact designed using Kirchhoff's law with hardly any mention of charge or Maxwell's equations.

It is obvious that low frequency, long time derivations of Kirchhoff's law are not able to deal with the modern use of electricity in the solid state devices of our digital technology. The derivations of Kirchhoff's law and discussion in textbooks and Wikipedia are unable to explain the successful use of the law to design circuits switching in $10^{-10}$ where $\partial\mathbf{E}/\partial t$ is not small, of the order of the power supply voltage per 100 psec, i.e., $5\ \times 10^{10}$ volts/sec.

This issue needs to be addressed, in my view, because the design of high speed circuits is the most used application of electrodynamics. It is embarrassing, to say the least, to use a derivation of the most widely used application of Maxwell's equations of electro***dynamics***—Kirchhoff's law—that is only true in the static case, when $\partial\mathbf{E}/\partial t = 0$.

**Other Applications.** Flux coupling by Kirchhoff's law is an important topic in the biophysics of ion channels and transporters. Flux coupling as described in the Appendix of *arXiv:1905.13574v3, 18 Jul 2019* has been used in several recent biological applications, as well as classical work discussed below.

Flux coupling has been exploited in reference [15] in a model of one of the most important membrane systems in biology, the cytochrome c oxidase system that helps generate ATP that is the source of chemical energy in almost all the chemical reactions of living metabolism, whether plants, animals or bacteria.

Reference [15] tries to present a complete model of cytochrome c oxidase both in its natural environment in a mitochondrion and in the artificial environment of a voltage clamped lipid bilayer. The properties of the two systems are quite different even though the protein cytochrome c oxidase is the same in both, just as the properties of ion channels are quite different in the natural setting and the voltage clamped apparatus. The model presented [15] does not need to keep track of the $\sim 10^{15}$ charges present in a mitochondrion because it uses Kirchhoff's current law to construct the model. Most published models use Coulomb's law (not even the Maxwell Gauss differential equation with its boundary conditions). It is not clear how these published models deal with the constraints imposed by the mitochondria on the current through cytochrome c oxidase. The total current (carried by all channels and transporters as well as the lipid membrane) sums to zero in such systems, because the potential in the interior of mitochondria is nearly uniform in space. In the jargon of biophysics, mitochondria are small compared to the length constant of biological cells.

The coupling of current in biological systems has also been used in models of water flow and current flow in the lens of the eye [13,16-26], and the optic nerve of the central nervous system of a salamander [13,27-31].

Of course, the coupling of current by Kirchhoff's law has been an ***implicit necessity in all models of nerve function*** since the work of Hodgkin [32] if not earlier. Davis [33] and Jack, Noble, and Tsien [34] develop these ideas and use more modern methods, as do many textbooks of biophysics, physiology, and their mathematical cousins, if not siblings.

In fact, Hodgkin [35] used the Kirchhoff coupling of sodium and potassium current (flowing through disjoint channel proteins far separated in space) in an important way. He used it to compare the currents in a voltage clamped axon, and a space clamped action potential (Fig. 10 of [35]), and thereby test the validity of the voltage clamp technique.

Voltage clamp had been introduced by Cole [36] and Marmont (working along side Cole in Cole's laboratory) and shown to Hodgkin by them [37-39]. This validation was in fact important historically in motivating the heroic computation of the action potential by Andrew Huxley that took many months of hand calculation, using a mechanical device that could add but not multiply. Electricity was not used to power the adding machines in Cambridge (UK) in those years. Computers were not available. Calculators were not available.

Hodgkin and his collaborators created modern electrophysiology and membrane biophysics (see Fig. 10 of [35]) and indeed transport biophysics, if one looks carefully into the study of active transport by Hodgkin and his associates in Physiology at Cambridge (UK). First they studied active transport in the squid nerve, and later in red blood cells.

The analysis of Hodgkin and coworkers, following [32], arose—I believe—from the circuit tradition developed by William Thomson[1] to design the trans-Atlantic cable that connected North America and Europe. The circuit tradition was brought to maturity by Heaviside [40-44].

Before the under the ocean cable, information took weeks to transfer across the Atlantic. After the cable, information was telegraphed in seconds. The British Empire celebrated the Atlantic cable as a triumph of British science and Imperialism well into the 1930s. The Empire included Cambridge (UK) Physiology.

**What if?** It is a rather frightening exercise in counter factual history to consider what would have happened if the circuit tradition had not been used to analyze the nerve action potential, if the Kirchhoff coupling of currents had been ignored, as it has been and is ignored in so many models and simulations in the chemical tradition.

How would molecular and (protein) structural biology, chemistry and biochemistry, have dealt with a propagating action potential? How would the action potential be computed from the $10^{15}$ charges involved using just Coulomb's law? How could the nerve signal be understood without Kirchhoff's law and the cable equation it implies? It seems unlikely that the thermodynamics or rate theory would have been successful.

It seems unlikely that any electrical properties of a system even as small as a mitochondrion can be calculated by dealing with charges, ignoring the conservation of total current. Nonetheless, that approach is still the bedrock for analysis and simulation in much of chemistry, biochemistry, and molecular biology, if not biophysics.

[1] Thomson has been more commonly known by his alias 'Lord Kelvin' after 1892 when he was renamed by the UK government as Baron Kelvin.

This counter factual example suggests that Kirchhoff's law should be given a prominent place in future models and simulations of proteins and nanodevices (*)*, a role comparable to its place in electrodynamics and circuit design. See Reference [123] of *arXiv:1905.13574v3, 18 Jul 2019* for a chemical perspective.

.

## **References in Addendum**

separate and independent from

References in the version arXiv:1905.13574v3*, 18 Jul 2019*,

December 4, 2022

**Bibliographical Note**.

The November 30, 2022 version of this paper is identical to the last revision

*arXiv:1905.13574v3, 18 Jul 2019*

except for the preceding section, and its references.

A different font is used for the new material

References in the two parts of this paper are distinct and independent and have different format.

Page numbering is also distinct and independent as is made clear by the different fonts and placement of page numbers,

I hope.

Version *arXiv:1905.13574v3, 18 Jul 2019 is on the following pages.*

# Kirchhoff's Current Law Can Be Exact

*available at https://arxiv.org/abs/1905.13574*

Robert S. Eisenberg
Department of Applied Mathematics
Illinois Institute of Technology;

Department of Physiology and Biophysics
Rush University Medical Center
Chicago IL

USA

[Bob.Eisenberg@gmail.com](mailto:Bob.Eisenberg@gmail.com)

*File name: Kirchhoff's Current Law Can Be Exact with Addendum December 5-2 2022.docx*

## Abstract

Kirchhoff's current law is thought to describe the translational movement of charged particles through resistors. But Kirchhoff's law is widely used to describe movements of current through resistors in high speed devices. Current at high frequencies/short times involves much more than the translation of particles. Transients abound. Augmentation of the resistors with *ad hoc* 'stray' capacitances is often used to introduce transients into models like those in real resistors. But augmentation hides the underlying problem, rather than solves it: the location, value and dielectric properties of the stray capacitances are not well determined. Here, we suggest a more general approach, that is well determined. If current is redefined as in Maxwell's equations, independent of the properties of dielectrics, Kirchhoff's law is exact and transients arise automatically without ambiguity. The transients in a particular real circuit—a high density integrated circuit for example—can then be described by measured constitutive equations together with Maxwell's equations without the introduction of arbitrary circuit elements.

Kirchhoff's current law says, in a crude representation, that the current that flows into a node, must flow out. In textbooks, Kirchhoff's current law describes the translational movement of charges through resistors that might be called the flux of electrons. The resistors are ideal, described by single real numbers. The current is carried by charges that have mass, e.g., electrons, and the movements are slow, without transients, nearly at DC [1-13].[2]

Kirchhoff's law is used today to describe currents through resistors on the nanosecond time scale. Indeed, it is the main design tool for the circuits of our high-speed technology. The currents in high speed circuits have transients not seen when movements are slow near DC. Current through resistors on the nanosecond time scale involves delays and overshoots: it is a complex phenomena [14-22], not just the movement of electrons in wires, in resistors or into capacitors. Kirchhoff's current law is viewed as approximate for these reasons, as is clear from its derivations [13, 14, 23, 24].

Engineers have dealt with these difficulties by *ad hoc* augmentation of DC circuits [25]. They construct wideband circuit models made of

(1) idealized resistors with current strictly equal $V_R/R$, where $R$ is the resistance and $V_R$ is the voltage across the resistor

in parallel with

(2) ideal capacitors, often characterized as stray [14, 26], sometimes as 'parasitic' [27].

These ideal capacitors carry currents strictly equal to $C\,\partial V/\partial t$, where $C$ is the capacitance, $V$ is the voltage across the capacitor and $t$ is time.

The size and location of the stray capacitances are chosen empirically so the augmented circuits more or less fit measurements of high-speed transients.

The values and locations of the stray capacitances are neither exact nor unique. They are often crude approximations, because actual currents deviate significantly from $C\,\partial V/\partial t$ or $V_R/R$ on the time scales of our digital technology, in ways important in practice [18, 19, 21, 22, 28-34].

[2] Sommerfeld [14], p. 101 describes the origin of Kirchhoff's laws, as Kirchhoff's solution to a problem posed in a seminar led by Neumann.

**Exact treatments have advantages.** Empirical and imprecise modifications of a circuit seem a poor substitute for an exact treatment, derivable from electrodynamics, if that is possible.

An exact treatment is possible using a rederivation of the law of conservation of current (eq. 3 below and ref. [25, 35-39]). We show (eq. 4, below) that Kirchhoff's law can be as exact as the Maxwell equations themselves[3], once current is defined as in the Maxwell equations, independent of the dielectric properties of matter [35-37, 40].

The role of 'current' was evidently a key issue in Maxwell's development of electrodynamics, according to the historical literature [41-44]. Maxwell defined current as we have, according to his successors at Trinity College Cambridge UK, Jeans and Whittaker, [45], p.511; [46], p. 280, respectively. Lorrain and Corson [1], p. 276 eq. 6-148, use that definition as well.

Current is usually defined as the flux of charge, although that is not the definition used here, see eq. (3)-(4) below. The flux of charge is not conserved. It accumulates in what are loosely called capacitors or stray capacitance. More precisely, the flux of charge accumulates according to the 'continuity equation' described in textbooks of electrodynamics for the oversimplified case in which single dielectric constant describes polarization [1, 2, 4, 5, 47].

Eq. 21-23 of ref [35] describes the effects of polarization on the accumulation of charge in general materials. Ref [35] derives the continuity equation for materials with complex polarization that cannot be described by a single dielectric constant.

**Current in the Maxwell equations.** Current appears in the equations of Maxwell in his generalization of Ampere's law.

**Maxwell's Version of Ampere's Law**

$$\frac{1}{\mu_o}\,\mathbf{curl\ B} = \mathbf{J} + \varepsilon_0\,\frac{\partial \mathbf{E}}{\partial t} \qquad (1)$$

$$\mathbf{J} = (\varepsilon_r - 1)\varepsilon_0\,\frac{\partial \mathbf{E}}{\partial t} + \mathbf{J}_{everything\ else} \qquad (2)$$

[3] Ref [19-23, 37, 41-43] apply the Maxwell equations within atoms using Bohm's version of quantum mechanics.

See texts [1, 2, 4, 5, 47] for the standard formulation of Maxwell's equations. See ref. [35] for an update to Maxwell's equation that includes a more realistic (and general) description of polarization and permanent charge (not present in Maxwell's original formulation). $\varepsilon_0\ \partial\mathbf{E}/\partial t$ is written separately in eq. (1) because it is a property of space, not matter, as discussed below. The variable $\varepsilon_r - 1$ and $\mathbf{J}_{everything\ else}$ are properties of matter, not space, see below on p. 6.

**B** describes the magnetic field with magnetic constant (permeability of vacuum) $\mu_o$. **E** describes the electric field, with electric constant $\varepsilon_0$ (permittivity of vacuum). $\varepsilon_r$ is the relative dielectric coefficient of perfect dielectrics, a single real positive constant $\geq 1$. $\mathbf{J}$ is the current produced by all translation of mass—including all movements of mass with charge, however small or transient the movement. $\mathbf{J}$ includes the polarization currents of dielectrics, ideal and real. The polarization of idealized dielectrics $(\varepsilon_r - 1)\varepsilon_0\ \partial\mathbf{E}/\partial t$ is isolated from other currents in eq. (2) only for convenience in relating our results to the literature [2, 5, 47-55]. Most of the literature of electrodynamics is written as if all dielectrics are ideal; Robinson [56] is a welcome exception that most resembles the treatment here.

Eq. (1)-(2) require a complete description of $\mathbf{J}$. A set of experimental results can serve this purpose. Theories or simulations of $\mathbf{J}$ can also serve this purpose if they fit the experimental data.[4]

$\mathbf{J}$ includes charge carried by the flux of particles, as in most textbooks. $\mathbf{J}_{everything\ else}$ includes the flux of particles and the nonideal polarization of real materials. It also includes other movements of charge, described below, and quantum effects [18-22, 36, 57-59] although those are not our focus.

**<u>Current</u>.** The current described as 'Everything else' in eq. (2) includes current produced by

(1) transport (flux) of electrons or charged particles, as in classical Kirchhoff's law of DC (or low frequency) circuit analysis.

'Everything else' also includes

(2) the nonideal properties of dielectrics;

---

[4] Those theories and simulations are most useful if they are transferable from one set of experimental conditions to another, using just one set of parameters. Not all theories and simulations have that property (see the note [61] and its documentation in [62]). In chemical kinetics, parameters are customarily adjusted as conditions change so a favored equation —the law of mass action—always fits data.

(3) polarization of matter in general, time dependent, nonlinear or whatever;

(4) currents driven by other fields, like diffusion, heat, and convection;

(5) any other movement (of any type) of charge with mass, including quantum effects [18-22, 36, 57-59].

'Everything else' does ***not*** include

(6) the properties of dielectrics idealized by $(\varepsilon_r - 1)\, \varepsilon_0\, \partial\mathbf{E}/\partial t$. That is treated separately as is the custom in the classical literature and textbooks.

(7) the polarization of the vacuum $\varepsilon_0\, \partial\mathbf{E}/\partial t$. That ethereal current is written separately from $\mathbf{J}$ and $\mathbf{J}_{everything\ else}$ in eq. (2).

**Polarization of the vacuum.** The term $\varepsilon_0\, \partial\mathbf{E}/\partial t$ is treated separately because it has such a different origin. $\varepsilon_0\, \partial\mathbf{E}/\partial t$ describes the 'polarization of the vacuum' [35] that allows—or should one say 'supports'?—the ethereal propagation of electromagnetic waves through a vacuum devoid of mass. The polarization of the vacuum is a consequence of general physical laws [2, 47, 49, 50, 60-62], not a property of matter, and should really be called 'the polarization of space'. The polarization $\varepsilon_0\, \partial\mathbf{E}/\partial t$ makes charge relativistically invariant, independent of velocity, even at velocities approaching the speed of light. References include p. 553 of [2] ; p. 228, eq. 5-110, of [1] and [47, 50, 60, 63, 64].

Charge is different from mass, length, and time, and most fundamental quantities. They vary with velocity according to the Lorentz transformation. Charge does not [2, 47, 49, 50, 60-62].

**Polarization of real materials.** Polarization of real materials is too complex to approximate usefully with a constant $\varepsilon_r$ [40]. Polarization is nonlinear in significant applications, particularly optical [56, 65-70]. Even when the polarization is linear, it involves complicated delays and varies too much with conditions and frequency/time to allow description with a constant $\varepsilon_r$.

An abundant and classical literature—prominent since at least 1928 [71, 72]—reports the actual properties of polarization of real dielectrics. Most of the classical literature concerns polarization that is proportional to the local electric field, i.e. that is linear [36, 71-100]. Polarization has been studied in great detail because it is a major determinant of the forces between molecules [101, 102].

Parsegian [101] discusses at length (and with admirable clarity) the connection between polarization and the spectra observed when light interacts with molecules. Spectra are used to identify molecules—more or less as successfully as fingerprints identify people—because spectra (and polarization) are sensitive to details of chemical structure and thus are remarkably diverse, almost as diverse as the molecules themselves [101, 103-107]. The diverse polarization and spectra of real materials obviously cannot be described by a single number or dielectric constant.

Despite this literature, $\varepsilon_r$ has been treated as a single real positive constant $\geq 1$ in textbooks of electrodynamics [1-5, 47, 49, 50, 61, 62] for many years apparently following [108, 109]. Robinson [56] is a welcome exception.

The properties of charge movement in matter are so complex that Feynman concluded that nothing much could be said in general (on p. 10-7 of [5]). It is necessary "… to exhibit in every case all the charges, whatever their origin, [so] the equations [of electrodynamics] are always correct." In fact, something important can be said about electrodynamics in general, independent of the properties of matter. Current, as defined by Maxwell, is universally conserved [35-39].

**<u>Conservation of current can be derived without reference to matter</u>:** Conservation of current $\mathbf{J}_{total}$ is in fact a general and exact property of the Maxwell equations, as general as the Maxwell equations themselves, independent of any properties of matter [35-39]. The dielectric constant $\varepsilon_r$ is not involved in the derivation of conservation of current at all. The derivation involves no statement or approximation to dielectric properties or polarization, nonlinear or linear, or any other properties of matter whatsoever.

**<u>Conservation of current</u>.** A general statement of conservation of current—eq. (3) below—can be derived [35-39] because the divergence of the curl in eq. (1) is identically zero, independent of any properties of matter, whenever Maxwell's equations can be used. The crucial term that produces universal conservation of current is the polarization of space $\varepsilon_0\ \partial\mathbf{E}/\partial t$. The polarization of space has nothing to do with matter because $\varepsilon_0$ has nothing to do with matter. $\varepsilon_0$ is a property of space, not matter.

Physically, the polarization of the vacuum creates the ethereal current $\varepsilon_0\ \partial\mathbf{E}/\partial t$. The ethereal current is an output of the Maxwell equations that varies so the total current $\mathbf{J}_{total} = \mathbf{J} + \varepsilon_0\ \partial\mathbf{E}/\partial t$ is conserved [35]. ***The ethereal current allows current to be conserved no matter what physics is involved in the translation of matter and charge***—see eq. 4 of [39] and Fig. 2 of [37]—even quantum physics (see eq. 45 of [36]).

*Conservation of Current*

$$\mathbf{div}\left(\overbrace{\mathbf{J} + \varepsilon_0 \frac{\partial \mathbf{E}}{\partial t}}^{\mathbf{J}_{total} = \boldsymbol{Current}}\right) = 0 \qquad (3)$$

**Novel derivation, apparently.** This general statement of conservation of current is not easily found in textbooks or the literature of electrodynamics.

The usual derivations of conservation of current (and the continuity equation derived in [35]) involve the dielectric constant $\varepsilon_r$ and treat it as a single real constant number in contradiction to experimental measurements of dielectric properties [36, 71-100], spectra [103-107, 110], and nonlinear polarization [56, 65-70], as mentioned before.

Derivations of conservation of current do not usually deal with the sum labelled *'Current'* in eq. (3). They do not usually deal with $\mathbf{J}_{total} = \mathbf{J} + \varepsilon_0\ \partial\mathbf{E}/\partial t$. The usual derivations deal with $\mathbf{J}$. Examination of eq. (3) shows that $\mathbf{J}$ is not universally conserved (because it neglects the ethereal current $\varepsilon_0\ \partial\mathbf{E}/\partial t$). $\mathbf{J}$ accumulates in systems, capacitors and stray capacitances as specified precisely by the continuity equation. $\mathbf{J}_{total}$ does not accumulate.

It is no wonder then that readers of the usual derivations, and most scientists, conclude (incorrectly) that conservation of current is a poor approximation that does not fit experimental data.

Conservation of total current is not an approximation when total current is defined as $\mathbf{J}_{total}$ because $\mathbf{J}_{total}$ includes the ethereal current $\varepsilon_0$, as shown in [35-39]. $\mathbf{J}_{total}$ is conserved. $\mathbf{J}$ is not.

**Kirchhoff's current law** is the general conservation law eq. (3) rewritten for branched one dimensional networks.[5] One dimensional networks have the special property that $\mathbf{curl\ B} = 0$, as is apparent if one writes out the curl operator explicitly in one dimension [111, 112]. Kirchoff's law then does not involve current flows or coupling induced by the magnetic field. Those phenomena occur in three dimensional problems and could be quite significant in rapidly changing signals, like those of modern integrated circuits if the circuits involved three dimensional flows of current. If the flows of current are one dimensional, phenomena mediated only by the magnetic field are not significant because $\mathbf{curl\ B} = 0$ in one dimensional circuits.

Kirchoff's law is widely viewed as an approximation, needing derivation; see p. 8-10 of [14] and [13, 23, 24] for some derivations, along with less precise discussions in most textbooks describing circuits. But Kirchhoff's law need not be approximate if current is defined[6] to include the ethereal term $\varepsilon_0\ \partial\mathbf{E}/\partial t$ [35-39].

[5] The precise definition of a network involves many issues beyond the scope of this paper [(6-12, 15-18, 114, 115].

*Proposed Definition*: $$\textbf{Current} \triangleq \mathbf{J}_{total} = \mathbf{J} + \varepsilon_0 \frac{\partial\mathbf{E}}{\partial t} \qquad (4)$$

With this definition of $\mathbf{J}_{total}$, all the current that flows into a node, flows out, exactly, at any time, no matter how brief, under all conditions in which the Maxwell equations apply. It is exactly conserved. It cannot accumulate at all. It cannot be stored. In the language of fluid dynamics, the electric current $\mathbf{J}_{total}$ is the flow of an (exactly and perfectly) incompressible fluid.

The incompressible fluids of hydrodynamics are, on the other hand, an approximation, and not a very precise one if one thinks of the dynamic range of electrical approximations (which are at least $10^7:1$ in real circuits and immeasurably large in systems with little matter, as in the space between stars). The incompressible flow $\mathbf{J}_{total}$ is not approximate.

$\mathbf{J}_{total}$ is as incompressible as Maxwell's equations are exact. It is perfectly incompressible under all conditions in which the ethereal current $\varepsilon_0\ \partial\mathbf{E}/\partial t$ is a perfect description of the physics of space, i.e. whenever Maxwell's equations are exact.

The question of what flow is described by $\mathbf{J}_{total}$ has received a great deal of attention from Maxwell and other workers, particularly in the special case of flow in a vacuum $\mathbf{J}_{total}(\textbf{vacuum}) = \varepsilon_0\ \partial\mathbf{E}/\partial t$ as described in Whittaker's "A History of the Theories of Aether & Electricity" [46]. The properties of the "luminiferous (a)ether" need not concern us as long as the equations describing the flow $\mathbf{J}_{total}(\textbf{vacuum})$ are correct, in the sense that they describe the properties of electromagnetic radiation in empty space, and mathematically consistent.

The question of what flows in the vacuum is too vacuous and too ethereal to grasp, in my opinion, because the crucial $\varepsilon_0\ \partial\mathbf{E}/\partial t$ term is a property of space, not matter. No one knows (as far as I can tell) how space can flow. From the point of view of mathematics, the meaning of the flow $\mathbf{J}_{total}(\text{vacuum})$ is determined by the properties of space and time, described by special and general relativity, as charge moves and the electric field evolves according to the Lorentz transformation [2, 47, 49, 50, 60-62].

[6] Lorrain and Corson [1], p. 276 eq. 6-148, use this definition of current. As far as I can tell, they do not discuss the approximate nature of $\varepsilon_r$. If the approximate nature of $\varepsilon_r$ is not discussed, the reader will then naturally think (incorrectly) that conservation of current is as unrealistic as the idealization of a single dielectric constant. The development of Robinson [57] does not depend on the approximate nature of $\varepsilon_r$, but as far as I can tell, it does not use the definition of current of eq. (4).

**Ordinary Definition of Current.** Current is defined in many textbooks as the charge carried by the translation of charged particles [2, 5-8, 11, 12, 14, 15, 17, 47-55, 113, 114]. This ordinary translational current is just part of the current $\mathbf{J}_{total}$ defined here in eq. (4). Another part of the current is classical, namely the current through an ideal dielectric $(\varepsilon_r - 1)\ \varepsilon_0\ \partial\mathbf{E}/\partial t$ found in eq. (2).

The current ordinarily defined in textbooks is not conserved. The continuity equation found in most textbooks of electrodynamics shows explicitly how charge accumulates so current (as ordinarily defined) is not conserved. In crude language, current can accumulate in capacitors and so is not conserved. Those capacitors include the 'capacitance of empty space' arising from $\varepsilon_0$. In more precise language, current accumulates in polarization according to the continuity equation written in various forms in eq. 21-23 of ref [35], appropriate for the complicated properties of polarization in real materials. The continuity equation for the flow of mass is of course also involved and must be included in a description of a coupled system.

We use Maxwell's definition of current $\mathbf{J}_{total}$ of eq. (4) precisely to avoid these difficulties. $\mathbf{J}_{total}$ does not accumulate, ever, anywhere. It is perfectly conserved. All the $\mathbf{J}_{total}$ that flows into a node flows out, always, everywhere.

The ordinary definition of current as only the flux of charged particles causes practical difficulty in Kirchhoff's law when applied to high speed circuits made only of idealized resistors (described by constant real numbers) [13, 14, 23-25]. Such idealized circuits do not have the transients, delays, or overshoots found in real circuits made of resistors [18, 23, 24, 28-34]. Idealized circuits cannot deal with the actual behavior of circuits observed on the nanosecond time scale if they use the ordinary statement of Kirchhoff's law and the ordinary definition of current as $\mathbf{J}$. Of course, the idealized circuits can be made more realistic by adding fictitious circuit elements.

**Stray Capacitances.** Engineers routinely add such fictitious capacitances to circuit models that include idealized resistors [25], so the transients of the idealized, augmented circuit approximate those observed in real circuits [26, 27]. The fictitious capacitances are not actual distinct circuit components. The location and values of the fictitious stray capacitances are chosen by the engineer to fit data and are rarely derived from electrodynamics.

The redefinition of current by equation (4) produces transients that are determined precisely without fictitious capacitances. Reference [25], eq. 14, shows how to choose capacitances arising from $\varepsilon_0\ \partial\mathbf{E}/\partial t$, once current is defined by eq. 4.

Transients in real circuits are more complicated than those predicted in reference [25]. Reference [41] shows how to include ***any*** types of current flow—polarization, inductive, or anything else—in the analysis, so the theory can cope with the properties of real circuit boards.

Polarization of real circuit boards involves more than $\varepsilon_r\varepsilon_0\ \partial\mathbf{E}/\partial t$ on the time scales of practical importance in our digital technology. Inductive effects of wires [20-24]; capacitances arising from materials like circuit boards [28-33]; and the complex geometry of the real circuit contribute additional terms beyond those arising from $\varepsilon_0\ \partial\mathbf{E}/\partial t$ or $(\varepsilon_r-1)\varepsilon_0\ \partial\mathbf{E}/\partial t$. The polarization in models of real circuits needs to be determined from measurements of real circuit boards because polarization is likely to depend on details of composition and construction of the real circuits, and perhaps on the signals themselves. Reference [41] shows how to include any polarization in the analysis, so it can describe real circuit boards.

**Role of the B field.** The lack of magnetic field **B** in branched one dimensional circuits is discussed at length in [35] where the speculation is made that the absence of **B** fields makes circuit design (at high speeds) much easier. **B** fields produce 'leaks of current' and cross talk that are difficult to deal with in circuit design, particularly because they are so variable and dependent on properties of the signals themselves, as well as details of layout, etc.

It seems to me that ground planes in high speed circuits might function better—i.e., more ideally, obeying the DC version of Kirchoff's law more accurately—if they too were built as branched one dimensional circuits, with minimal **B** fields, cross talk, and current leakage in the grounds.

**Conclusion.** Stray capacitances, of unknown value, location, and unrealistic properties seem a poor substitute for currents defined exactly by the Maxwell equations. Transients arise naturally if current in Kirchhoff's law is defined as in Maxwell's equation, see eq. (1)-(4). Realistic transients, arising from nonideal properties of circuits, can be easily incorporated into our treatment [35].

Appendix

**Networks of Chemical Reactions**

Our discussion has focused on Kirchhoff's law in electrical networks. Networks and analogs of Kirchhoff's law are also used widely to describe interacting chemical reactions in chemistry, biochemistry and biology [115-122]. The hundreds of enzyme reactions in the intermediary metabolism that form 'the chemical factory of life' are a notable example. These networks of chemical reactions are used by thousands of scientists every day to explain medical and biological phenomena and appear in every textbook of physiology, biochemistry, molecular and cell biology, and medicine.

**Chemical networks.** Equations of chemical [115-120] and enzyme kinetics [121, 122] use analogs of Kirchhoff's law to connect chemical reactions. The analogs describe the flux of particles, not the flow of electric current. The analog equations use conservation of mass but rarely mention or use conservation of current. They do not include an $\varepsilon_r \varepsilon_0\ \partial\mathbf{E}/\partial t$ term and so cannot conserve $\mathbf{J}_{total}$. They also do not conserve $\mathbf{J}_{total}$ in the steady-state, as is clear by using the law of mass action of chemical reactions to compute and compare steady state currents for reactions in series [35, 37, 123, 124].

**Auxiliary conditions may be possible** that make equal the current $\mathbf{J}_{total}$ in a series of chemical reactions but auxiliary conditions are not found in the standard references [115-122] or textbooks, simple or advanced, as far as I know. The implication is that the treatment of chemical reactions in a wide range of the chemical and biological literature is incompatible with the Maxwell equations. One consequence may be the well known need to adjust parameters of chemical reactions as reactions are transferred from one set of conditions to another. Descriptions of chemical reactions are expected to be transferable only if they describe the real world properties of the reactants [35, 37, 123, 124], including its electrodynamics.

The real world is different from the world of textbooks, or chemical reaction theory because the real world follows Maxwell's equations universally, at all locations and times, and under all conditions in which chemical reactions are studied. In the real world, the atomic scale electric field in one reaction is influenced by distant reactions—and by even more distant macroscopic boundary conditions—to create 'flux coupling' that would not be present without this long range influence of the electric field. Rate 'constants' of a series of reactions are coupled even for chemical reactions that are distinct and disjoint in space, far apart on an atomic length scale, occurring in different structures.

The auxiliary equations needed to conserve current $\mathbf{J}_{total}$ interact with the equations of chemical kinetics that conserve mass. One set of conservation laws is not enough. Both must be solved together. One way to do this is to write separate networks for electric current and for mass flow.

The networks for conservation of mass and conservation of current $\mathbf{J}_{total}$ together can be solved with the methods of the theory of complex fluids which are designed to deal with multiple simultaneous force and flow fields. In particular, the variational method of complex fluids [125-136] may allow one to create a new synthetic composite functional with units of energy. In this way, flux coupling can be dealt with consistently on all scales, i.e., with all variables satisfying all boundary conditions with one set of unchanging parameters.

**Flux coupling** plays a central role in many transport systems and chemical reactions in biology. Flux coupling allows the 'unnatural' uphill transport of ions (and solutes) to be driven by the natural downhill movement of other ions (and solutes). Active transport of this sort occurs throughout biological cells and organelles and is one of the fundamental mechanisms of life.

Flux coupling is a central mechanism in oxidative phosphorylation and photosynthesis, which are the ultimate source of the energy for life, whether the energy is used in chemical, electrical, or diffusional processes. In oxidative phosphorylation and photosynthesis, electron flow is coupled to the movement of ions and 'protons' across membranes of mitochondria.[7] Flux coupling in oxidative phosphorylation, and in photosynthesis, are particular examples of a general phenomenon. Flux coupling comes from a combination of macroscopic and atomic scale phenomena ***involving their mutually generated electric field,*** as well as (perhaps) chemical interactions.

**Flux coupling depends on the experimental conditions.** Flux coupling depends on anything that can change the macroscopic electric field. In particular, flux coupling depends on the boundary conditions.

Boundary conditions are different in different setups used to measure flux and flux coupling. Some setups leave the system in its natural state, with transporters/channels in native structures, like mitochondria. Other setups insert the transporter/channel into a lipid bilayer.

Flux coupling will be different in the two setups because they impose different boundary conditions. In mitochondria, the sum of currents across a membrane is zero (as it is in cells or organelles shorter than a length constant or so [137]). The biological situation imposes the condition that the sum of all currents across the membrane is zero, because structures, like mitochondria, are so small.

---

[7] 'Proton' is a nickname for the positively charged form of water, sometimes written as $H_3O^+$.

Transporters are often studied, however, in non-native bilayer setups where currents are only constrained by the applied electrical and chemical potential (i.e., concentrations). Quite different results for flux coupling can occur in that case, quite different from the flux coupling that is found in the mitochondria, because the constraints on the fluxes (the boundary conditions produced by conservation of current) are so different.

**Flux Coupling and Channel Opening.** Manuel Landstorfer and I recently realized [36] that a different kind of coupling can occur between the voltage sensor of a voltage sensitive channel [138-143] and its conduction pore. Part of the current injected by the voltage sensor might flow through the conduction pore and trigger its opening.

Similarly, flux coupling can occur when a transmitter binds to a receptor on a channel protein, like acetylcholine binding to the acetylcholine channel. Binding of a charged agonist to a charged receptor produces current and some of that current can be injected into the conduction pore, triggering its opening, even though the pore is far away from the receptor.

**Historical Note.** The reason $\mathbf{J}_{total}$ is not conserved in chemical kinetics [115-122] is historical, I suspect. Chemists were understandably focused on mass and its transformations, not electric current and its flow. Conservation of mass is used in the derivation of the equations of enzyme kinetics, but conservation of current is not mentioned, to the best of my knowledge.

Mass can accumulate according to the continuity equation of fluid dynamics. Chemical kinetics allows accumulation of mass this way but is silent about conservation of charge or current, whether steady-state or transient. Charge can accumulate according to the continuity equation for $\mathbf{J}$ shown in various forms in eq. 21-23 of [35]. In contrast, the Maxwell equations do not allow accumulation of the current $\mathbf{J}_{total}$, whether steady-state or transient, not at all, not under any conditions or at any time. All the $\mathbf{J}_{total}$ that flows into a node flows out.

**Conclusion.** It seems wise to use network models in both chemistry and electronics that conserve $\mathbf{J}_{total}$ with as few *ad hoc* extensions to Maxwell's equations as possible: the artifacts in electrical potential and fluxes can be very large if current is not conserved. The artifact in the electric potential is large, because of the strength of the electric field (see the first page of Feynman [5] and Appendix of [37]). Fluxes often flow over large barriers, where flux is an exponential function of potential. Flux artifacts can then be exponentially large and are best avoided. Models with large artifacts are unlikely to be transferrable from one of conditions to another.